\documentclass[11pt,a4paper]{article}
\usepackage[utf8]{inputenc}
\usepackage{graphicx,verbatim,array,multicol,courier}
\usepackage{amsmath,amssymb,amsthm,mathrsfs, mathtools}
\usepackage{color, graphics, graphicx, siunitx}
\usepackage{amsfonts, dsfont, bm}
\usepackage{natbib} 
\usepackage{a rydshln}
\usepackage{geometry}
\usepackage{enumitem,float}
\usepackage{lscape}
\usepackage[pdftex,bookmarks,colorlinks]{hyperref}
\usepackage[dvipsnames]{xcolor}
\usepackage{tabularx}
\usepackage{siunitx}
\usepackage[labelformat=simple]{subfig}
\usepackage[linesnumbered,ruled,vlined]{algorithm2e}
\usepackage{booktabs}

\newlist{inparaenum}{enumerate}{2}% allow two levels of nesting in an enumerate-like environment
\setlist[inparaenum,1]{label=(\alph*)}% labels for top level
\setlist[inparaenum,2]{label=(\roman{inparaenumi}\emph{\alph*})}% labels for second level

\makeatletter
\def\adl@drawiv#1#2#3{%
        \hskip.5\tabcolsep
        \xleaders#3{#2.5\@tempdimb #1{1}#2.5\@tempdimb}%
                #2\z@ plus1fil minus1fil\relax
        \hskip.5\tabcolsep}
\newcommand{\cdashlinelr}[1]{%
  \noalign{\vskip\aboverulesep
           \global\let\@dashdrawstore\adl@draw
           \global\let\adl@draw\adl@drawiv}
  \cdashline{#1}
  \noalign{\global\let\adl@draw\@dashdrawstore
           \vskip\belowrulesep}}
\makeatother

\numberwithin{equation}{section}
\theoremstyle{definition}
\newtheorem{defi}{Definition}[section]
\newtheorem{cond}[defi]{Condition}
\theoremstyle{plain}

\newtheorem{prop}[defi]{Proposition}

\newtheorem{cor}[defi]{Corollary}

\theoremstyle{remark}

\theoremstyle{example}
\newtheorem{ex}[defi]{Example}

\newcommand{\diff}{\mathrm{d}}

\definecolor{navy}{rgb}{0,0,0.502}
\definecolor{brown}{rgb}{0.59, 0.29, 0.0}
\hypersetup{colorlinks,%
	citecolor=blue,%
	filecolor=green,%
	linkcolor=red,%
	urlcolor=violet,%
}

\newcommand{\hell}{{\mathscr{H}}}

\newcommand{\Real}{\mathbb{R}}

\newcommand{\Prob}{\mathbb{P}}

\newcommand{\bftheta}{{\boldsymbol{\theta}}}

\title{Statistical Prediction of Peaks Over a Threshold}
\author{S. A. Padoan\\
	Department of Decision Sciences, Bocconi University, Italy\\
	and\\ 
	S. Rizzelli\\
	Department of Statistical Sciences, University of Padova, Italy
}
\begin{document}

\maketitle
\begin{abstract}
In many applied fields it is desired to make predictions with the aim of assessing the plausibility of more severe events than those already recorded to safeguard against calamities that have not yet occurred. This problem can be analysed using extreme value theory.
%is an extreme values theory problem. 
%We consider the popular peaks over a threshold method and  that the generalised Pareto approximation of density functions of a future unobservable excess and peak random variables can be very accurate. 
We consider the popular peaks over a threshold method and show that the generalised Pareto approximation of the true predictive densities of both a future unobservable excess or peak random variable can be very accurate. We propose both a frequentist and a Bayesian approach for the estimation of such predictive densities. We show the asymptotic accuracy of the corresponding estimators and, more importantly, prove that the resulting predictive inference is asymptotically reliable. We show the utility of the 
%predictive based inference 
proposed predictive tools
%the 
analysing  extreme temperatures in Milan in Italy.
\end{abstract}

{\it Keywords:}  Contraction rate, Exceedances, Extreme index, Generalised Pareto, Predictive density, Probabilistic forecasting.

%%%%%%%%%%%%%%%%%%%%%%%%%%%%%%%%%%%%%%%%%%%%%%%
%
\section{Introduction}
%
%%%%%%%%%%%%%%%%%%%%%%%%%%%%%%%%%%%%%%%%%%%%%%%

The prediction of a value that a random variable can plausibly assume in the future is one of the most important statistical problems. In many applied fields as in finance, economy, climate science, physics, etc., it is desired to make predictions with the aim of assessing the plausibility of  more severe events than those already recorded, in order to safeguard against calamities that have not yet occurred, as for instance  a global financial crisis, heavy monetary losses due to catastrophic weather, etc. This is an extreme value %theory 
problem and can be analysed using Extreme Value Theory
 \citep[e.g.][]{beirlant2006, dehaan+f06, resnick2007, falk2010}. 

The Extreme Value Theory (EVT) provides probabilistic models and, based on them, statistical methods to assess the risk related to future extreme episodes.
% still unobserved yet. 
However, the majority of the discussed methods focus on estimation %techniques 
of a distribution parameter, as for instance the quantile corresponding to an infinitesimal exceeding probability or the right end-point of the true data distribution. 
These quantities are representative of future extreme, yet unobserved values for the phenomenon under study and, to a certain extent, can be regarded as point forecasts.
% These quantities are representative of large, yet unobserved values for a phenomenon under study and, to a certain extent, can be regarded as point forecasts of a future extreme value.
%
The probabilitstic forecasting method 
is unarguably a richer approach 
%for the 
to prediction, as it allows to recover the entire distribution or density of an unobserved random variable and accordingly better assess the uncertainty of future events. 
A widely used tool for probabilistic based predictions are the predictive densities \citep[e.g.,][]{gneit07JASA} or density forecasts \citep[e.g.,][]{gneiting14}. 
%When probability density
%When probabilistic predictions take the form of a probability density, they are typically referred to as predictive densities \citep[e.g.,][]{gneit07JASA} or density forecasts \citep[e.g.,][]{gneiting14}.  

In this paper we focus on the Peaks Over a Threshold (POT) method, which is one of the most popular approaches for modelling univariate extremes \citep[e.g.][]{balkema1974residual, davison1990}. We begin specifying sufficient conditions under which we can show that a Generalised Pareto (GP) density 
%is
provides
a good approximation of the density of a suitably normalised excess variable $X-t|X>t$, for the threshold $t$ that goes to the right end-point of the distribution of $X$,
%infinity,
 in Hellinger distance \citep[see, e.g.,][for the notion of Hellinger distance]{vdv2000}.
Then, given a sample of independent and identically distributed (i.i.d.) random variables $X_1, \ldots, X_n$ whose common distribution $F$ satisfies the above conditions,
% \citep[][Ch 1]{dehaan+f06}, 
we consider both the prediction of a future unobservable excess random variable $(X_{n+1}-Q(p))|X_{n+1}>Q(p)$ and a peak random variable $X_{n+1}|X_{n+1}>Q(p)$, for different probability levels $p$ and corresponding
 ($1-p$)-quantiles $Q(p)$. We allow that  $Q(p)$ 
 is an increasingly high threshold by setting 
the exceeding probability as $p=p_n$, with $p\to0$ as $n\to\infty$. In particular, we consider both the so-called {\it intermediate} case where $p=k/n$, with  $k=k_n$, 
%with 
$k\to\infty$ as $n\to\infty$ and $k=o(n)$, and the so-called {\it extreme} case where 
$p\ll k/n$, with $np\to a>0$ as $n\to\infty$.
Exploiting EVT and in particular the threshold stability property  
\citep[e.g.][Ch 1, 5]{falk2010} we consider a class of GP approximations of the true unknown predictive density functions, and show they are very accurate in Hellinger distance.
We propose a frequentist and a Bayesian method for estimating
the predictive densities of both future unobservable excess and peak variables and we prove that the estimated densities are close to the true densities in Helliger distance, with probability tending to one. 
More importantly, when an extreme region is defined as a subset of the support of the estimated peaks'  predictive densities with a corresponding predictive probability level $1-\alpha$, then we prove its coverage probability (under the true peaks-generating distribution) reaches the nominal level $1-\alpha$ asymptotically.

Finally, we analyse the extreme temperatures over the last thirty years in Milan (Italy), where in 2003 the record temperature of 38.3°C has been recorded, due to a vast heatwave. We explain how to
% estimate the predictive density
derive a predictive density of a future  
peak temperature for different values of a high threshold and in particular how to suitably  derive
%derive 
very high thresholds. We %derive 
specify what are the thresholds beyond which a future peak can plausibly reach the record temperature in 2003 or being even hotter. We can then understand how much a peak temperature can plausibly rise beyond the temperatures observed so far.

The online supplementary materials contains technical results and proofs. Section \ref{sec:background} of the article provides a short EVT  background and our first result on the GP density approximation. In Section \ref{sec:strong_results} we
consider the GP approximation of the true predictive densities and 
discuss their accuracy, on their basis we
propose probabilistic forecasting tools
and show the reliability of the corresponding inference. 
%The estimators are applied
We illustrate the use of these procedures analysing a real temperature dataset in Section \ref{sec:real_analysiss}. The article ends with a brief discussion on future research prospects in Section \ref{sec:conclusion}.
%
 %the Hellinger distance, defined as 
%	\[
%	\hell(p,q)=\left[
%	\int_{\mathcal{X}} \left(
%	\sqrt{p(x)}-\sqrt{q(x)}
%	\right)^2 \diff \nu(x)
%	\right]^{1/2}
%	\]
%	for any two probability densities $p,q$ with respect to a measure $\nu$ over a space $\mathcal{X}$.

%%%%%%%%%%%%%%%%%%%%%%%%%%%%%%%%%%%%%%%%%%%%%%%
%
\section{Background}\label{sec:background}
%%%%%%%%%%%%%%%%%%%%%%%%%%%%

Let $X$ be a random variable with an unknown distribution $F$, whose right end-point is denoted by $x^*$. Extreme value methods rely on the elementary assumption that $F$ is in the {\it domain of attraction} of an extreme value distribution \citep[][Ch. 1.2]{dehaan+f06}. This feature can be simply formulated as follows. 
Consider the distribution of the {\it excess} variable $X-t|X>t$,
\begin{equation}\label{eq:ConDist}
F_t (x) = \Prob(X-t \leq x | X>t)= \frac{F(x+t)-F(t)}{1-F(t)}, \quad t<x^*, x >0.
\end{equation}
If there is a scaling function $s(t)>0$, with $t< x^*$, such that
\begin{equation}\label{eq:GP}
\lim_{t\to x^{*}}F_t(s(t)x)=H_\gamma(x):=
\begin{cases}
1-\left(1+\gamma x\right)^{-1/\gamma},\quad x \in \mathcal{S}_\gamma,
\\0, \hspace{8em}
\text{otherwise},
\end{cases}
\end{equation}
for all continuity points $x$ of $H_\gamma(x)$, then we say that $F$ is in the domain of attraction of $H_\gamma$, 
%and we shortly denoted it as $F\in\DoA(H_\gamma)$, 
see, e.g., \citet[][Theorem 1.2.5]{balkema1974residual, dehaan+f06}. Specifically, $H_\gamma(x)$ is named 
standard Generalised Pareto (GP) distribution \citep[e.g.,][Ch. 2.2]{falk2010}, where $\gamma\in\Real$ is the so-called {\it extreme value index}, which describes the right-tail heaviness of $F$, and  $\mathcal{S}_\gamma=(0,\infty)$ if $\gamma\geq 0$ and $\mathcal{S}_\gamma=(0,-1/\gamma)$ otherwise.
Its density function is
$$
h_\gamma(x)=\left(1+\gamma x \right)^{-(1/\gamma+1)},\quad x\in \mathcal{S}_\gamma.
$$ 
%
%{\color{magenta}From a distribution viewpoint} 
%{\color{red}More generally,}
The GP family of distributions is in general a two-parameters class of the form $H_{\bftheta}(x):=H_\gamma(x/\sigma)$, with $\bftheta=(\sigma,\gamma)$ and where $\sigma>0$ is a scale parameter. Its density is $h_{\bftheta}(x):=h_\gamma(x/\sigma)/\sigma$ with $x\in\mathcal{S}_\bftheta$, with $\mathcal{S}_\bftheta=(0,\infty)$ if $\gamma\geq 0$ and $\mathcal{S}_\bftheta=(0,-\sigma/\gamma)$ otherwise. 

The practical utility of the asymptotic result in \eqref{eq:GP} is that although $F$ is unknown,  as soon as $t$ and $y$ are large enough values (e.g. $y\equiv y_t=s(t)x$), then the distribution $F_t(y)$ of the excess $X-t|X>t$ can be approximated by the GP distribution $H_{\gamma}(y/s(t))$.  The {\it threshold stability} is a key property of the GP family. Precisely, GP distributions are threshold stable, which means that if $Y$ follows a GP distribution $H_\bftheta$ and if for $u\geq0$ we have $H_\bftheta(u)<1$ and $\sigma+\gamma u>0$, then $Y-u | Y>u$ follows the GP distribution $H_{\bftheta'}$, with $\bftheta'=(\sigma+\gamma u, \gamma)$. In other words, the thresholding operation does not change the original distribution apart from a scaling factor. Thus, if another threshold $t'>t$ is selected, then the distribution of the excess $X-t'|X>t'$ can still be approximated by the GP distribution $H_{\gamma}(y/s(t'))$, where only the scale  parameter $s(t')=s(t)+\gamma(t'-t)$ is changed. This is practically relevant in applications as one can progressively increase the threshold and still use the same GP distribution, but with the suitable scaling parameter, 
%as basis 
in order to extrapolate extreme events further and further into the tail of the distribution.

In the sequel, we assume that $F$ allows for a density function and we denote by $f_t$ the density of the conditional distribution $F_t$. In real applications working with the {\it peak} variable $X|X>t$ allows for a more intuitive interpretation of
what an extreme event is. We then also consider for 
$y=x+t$ 
the distribution of the peak
$$
G_t(y)=\Prob(X\leq y | X>t)=
\begin{cases}
	 \frac{F(y)-F(t)}{1-F(t)}, \quad y > t,\\
	 0, \hspace{4.5em} \text{otherwise}
\end{cases}
$$
and its density $g_t(y)=f_t(y-t)$. In this case, the distribution $G_t(y)$ is approximated by $H_{\gamma}((y-t)/s(t))$.

For $0\leq p \leq 1$, let $Q(p):=F^{\leftarrow}(1-p)$, where $F^{\leftarrow}$ is the left-continuous inverse function of $F$. The domain of attraction condition in  in formula \eqref{eq:GP}  can be equivalently stated 
in terms of the following convergence result
\begin{equation}\label{eq:quant}
\lim_{v\to\infty}\frac{Q(1/(vx))-Q(1/v)}{a(v)}=\frac{x^\gamma-1}{\gamma},
\end{equation}
for $v\geq 1$ and all $x>0$, where $a(v)>0$ is a suitable scaling function. In particular, we have that $s(t)= a(1/(1-F(t)))$, see \citet[][Ch. 1]{dehaan+f06} %Theorem 1.1.6 and 
for suitable selections of the function $a$.

Stronger convergence forms than that in formula \eqref{eq:GP} have been established under further conditions at the density level \citep[see e.g.][]{raoult03, padoan24}. In particular, the convergence result in Hellinger distance of \cite{padoan24} is especially useful for statistical purposes when performing probabilistic forecasting. It has been derived under the following second order von Mises-type condition \citep[see][Condtion 4]{deHaan96, raoult03}
\begin{cond}\label{cond:SO}
The distribution function $F$ is twice differentiable. The function 
%and there exists $\rho\leq 0$ such that
%
$$
A(v):= \frac{vQ''(1/v)}{Q'(1/v)}+1-\gamma
$$
is such that is of constant sign near infinity and its absolute value $|A(v)|$ is regularly varying as  $v\to\infty$ with index of variation  $\rho\leq 0$.
\end{cond}
Let $l_t( \, \cdot \, )= f_t( s(t) \, \cdot \, ) s(t)$ denote the density function of $(X-t)/s(t)|X>t$, with $s(t)=(1-F(t))/ F'(t)$.
Under Condition \ref{cond:SO}, with  $v=1/(1-F(t))$, \cite{padoan24} established that for a generic $\gamma \geq 0$
the density $l_t$  converges to the density $h_\gamma$ in Hellinger distance and the speed of convergence is $|A(v)|$.

The interest in this article is to extend such result to the case that
 $\gamma>-1/2$, which is typically considered when using 
important inferential tools as the maximum likelihood method and Bayesian approach 
to analyse block maxima \citep{b+s2017, dombry19, padoan24b} and peaks over threshold \citep{drees04, dehaan+f06, dombry23}. In particular, in this case the GP family satisfies some regularity conditions useful for the likelihood based inference, see \citet{dombry23}. Therefore, we provide here the following approximation result.
\begin{prop}\label{theo:hellrate}
	Assume Condition \ref{cond:SO} is satisfied with $\gamma >-1/2$. Then, there exist constants $0<c<C<\infty$ and $t^*< x^*$ such that, for all $t \geq t^*$,
	\begin{equation}\label{eq:hellrate}
		c |A(v)| \leq \hell(l_t, h_\gamma ) \leq C |A(v)|{.}
	\end{equation}
	%
	%, where $v=1/(1-F(t))$ and $l_t( \, \cdot \, )= f_t( s(t) \, \cdot \, ) s(t)$.	
\end{prop}
As shown in the sequel, the key statistical benefit of the above probabilistic result is to allow to control the bias amount in density estimation for the peaks over threshold method and assess the accuracy of the corresponding probabilistic forecasting method for future extremes.

%%%%%%%%%%%%%%%%%%%%%%%%%%%
\section{Probabilistic forecasting}\label{sec:strong_results}
%%%%%%%%%%%%%%%%%%%%%%%%%%%

In the statistical field, predicting
a future value not yet observed, on the basis of an available data sample, is one of the most practically relevant tasks in applications. This is an ambitious mission within the extreme values context, as the target is studying the occurrence of 
a future event that is even more extreme than those that have been observed in the past and that can
have a strong impact in society real life. A powerful prediction approach in statistics is the so-called probabilistic forecasting, that aims to derive the entire predictive distribution or density function of a future unobservable variable. 
For instance, the predictive density can be used for the construction of reliable predictive regions. Probabilistic forecasting for extreme events is not an ordinary statistical task, as it must be based on asymptotic results (e.g., that in \eqref{eq:GP} or other forms of convergence). The goal of this section is to show that the GP densities provide accurate approximations of the true densities of excesses and peak variables and 
that probabilistic forecasting techniques based on them 
are thus reliable.

Assume that a sample $X_1,\ldots,X_n$ of i.i.d. random variables with a common unknown specific distribution $F_0$ is available. In the sequel, to simplify the notation we do not report the subscript ``$0$'' in true distributions, densities, parameters, etc., when its is clear from the context that we refer to the true quantities, while we report it when strictly necessary. In the EVT, a typical approach to select a high threshold is to consider an intermediate sequence $k=k_n$, satisfying $k\to \infty$ as $n\to\infty$ and $k=o(n)$, and then use 
the tail quantile function to define the threshold $t=Q(k/n)$. Let $(X_{n+1}-t)|X_{n+1}>t$ be a future unobservable excess variable, which we assume independent from the excesses arising from the original sample $X_1,\ldots,X_n$. Then, under Condition \ref{cond:SO}, its true predictive 
density $f^*_t$ can be approximated by $h_{\gamma}(\cdot/s(t))$, provided that $t$ is large enough. Similarly, let $X_{n+1}|X_{n+1}>t$ be a future unobservable peak variable. Then, its true predictive density $g^*_t$ can be approximated by $h_{\gamma}((\cdot-t)/s(t))$.
Let $t'=Q(p)$ be a further threshold even larger than $t$, with $p\equiv p_n$ 
satisfying $p\to0$ as $n\to\infty$ and $p\leq k/n$. 
In the sequel we allow for different speeds and in particular for the case when
$np\to a>0$ as $n\to\infty$, i.e. when the expected number of exceedances approaches a positive constant, then we recall that in this case $Q(p)$ is called an extreme quantile \citep[see, e.g.,][Ch. 4]{dehaan+f06} and $p$ is the so-called extreme level \citep[e.g.,][Section 2.2]{p+s22}.
By the threshold stability property, the predictive density of a future unobservable excess variable $f^*_{t'}$ can be approximated by
	$$
	%X^*\stackrel{d}{=}(X_{n+1}-t')|X_{n+1}>t',\quad\text{by}\quad
	f_{p}^*(x):= \frac{h_{\gamma} \left(
		x /({\left(np/k\right)^{-\gamma} s(t))}
		\right)}{\left(np/k\right)^{-\gamma} s(t)}
	$$
	and that of a future unobservable peak variable $g^*_{t'}$  by
	$$
	%X^*\stackrel{d}{=}X_{n+1}|X_{n+1}>t', \quad\text{by}\quad
	g_{p}^*(x):=f_{p}^*\left(
	x- t-s(t) \frac{\left(n p / k\right)^{-\gamma}-1}{\gamma}  \right).
	$$
Leveraging on the results of Proposition \ref{theo:hellrate}, next corollary rigorously establishes that the densities $f_{p}^*$ and $g_{p}^*$ provide accurate approximations of the true densities $f^*_{t'}$ and $g^*_{t'}$, over a range of possible 
small exceedance probability levels
$p\leq k/n$.
\begin{cor}\label{cor:newapprox}
Assume that Condition \ref{cond:SO} is satisfied with  $\gamma>-1/2$. 
Let $t=Q(k/n)$ and $t'=Q(p)$. 
\begin{inparaenum}
			\item[(i)] If $ k/(np) \to \nu  \in [1,\infty)  $ as $n \to \infty$, we have
			$$
				\hell(f_{t'}^*, f_p^*)=O\left(|A(n/k)| \right), \quad \text{and}\quad
			 \hell(g_{t'}^*, g_{p}^*)=O\left(\sqrt{|A(n/k)|} \right).
			$$
\item[(ii)] If $k/(np) \to \infty$ and $-A(n/k) \log(np/k)\to 0$ as $n\to\infty$, we have
	$$
			\hell(f_{t'}^*, f_p^*)=O\left( 
			-|A(n/k)| \log(np/k)
			\right).
	$$
		    \item[(iii)] If $k/(np) \to \infty$ and  
		    $
		    A(n/k) w_{\gamma}\left(k/(np)\right) \to 0 
		    $
		    as $n \to \infty$,
		    with
		    $$
		    w_{\gamma}(x) := \begin{cases}
		    	\log(x), \hspace{1.3em} \gamma>0,\\
		    	\log^2(x), \quad \gamma=0,\\
		    	x^{-\gamma}, \hspace{2.4em} \gamma<0,
		    \end{cases} 
		    $$
		    where $x>0$, we have
			$
			\hell(g_{t'}^*, g_{p}^*)=O\left( \sqrt{A(n/k)w_{\gamma} \left(k/(np)\right)}\right).
			$
\end{inparaenum}
\end{cor}
Probabilistic forecasting based on the
predictive densities can be achieved by estimating 
the densities $f_{p}^*$ and $g_{p}^*$ for a range of different extreme levels $p\leq k/n$.  A possible way to do it
is as follows. First of all, we specify a high threshold $t=Q(k/n)$ setting an exceeding probability equal to  $k/n$, the so-called effective sample fraction, whose value falls (on  average) within the sample and can be easily  
estimated by the $(n-k)$th order statistic $X_{n-k,n}$. Secondly, we observe that, for different values of $p$, the densities $f_{p}^*$ and $g_{p}^*$ can be 
inferred through the parametric GP densities
\begin{align}\label{eq:predictive_dist}
\nonumber f^{*}_{\bftheta,p}(x)&:= \frac{h_{\gamma} \left(
		x /({\left(np/k\right)^{-\gamma} \sigma)}
		\right)}{\left(np/k\right)^{-\gamma} \sigma},\\
	g^*_{\bftheta,p}(x)&:= f^*_{\bftheta, p}\left(x-X_{n-k,n} - \sigma \frac{\left(\frac{np}{k}\right)^{-\gamma}-1}{\gamma}\right),
\end{align}
where $\bftheta=(\sigma,\gamma)$, with $\sigma>0$ that represents the scaling function $s(t)$ for a given fixed high threshold $t$ and $\gamma>-1/2$. Then, the $k$th larger order statistics $X_{n-k+1,n},\ldots,X_{n,n}$ can be used to carry out inference on
the parameter $\bftheta$, which 
allows in turn the estimation of such two
predictive densities, simultaneously for all the desired values of $p$.
In the next subsections we describe a frequentist
and a Bayesian 
approach to the estimation of $f^*_{\bftheta, p}(x)$ and $g^*_{\bftheta,p}(x)$, we establish the consistency and contraction rates of the proposed estimation methods and finally discuss the implications that these results have in probabilistic forecasting terms.

\subsection{Frequentist approach}\label{sec:freq}
A simple frequentist approach for the estimation of 
the densities $f^*_{t'}$ and $g^*_{t'}$ of excess and peak variables corresponding to exceeding probability $p\leq k/n$, is as follows.
Let $T_{n,i}$, $i=1,2$ be suitable measurable functions. Let 
$$
\widehat\gamma_n=T_{n,1}(X_{n-k,n},..., X_{n,n})
$$ 
be a generic sample estimator of the tail index $\gamma>-1/2$ and 
$$
%\widehat\gamma_k=T_{k,1}(X_{(n-k+1)},..., X_{(n)}) \quad\text{and}\quad
\widehat{\sigma}_n=T_{n,2}(X_{n-k,n},..., X_{n,n})
$$ 
be a generic sample estimator of the scaling parameter $\sigma>0$. Exploiting the GP density approximations {of} $f^*_{t'}$ and $g^*_{t'}$, plug-in estimators of them are simply obtained by
$$
{\widehat{f}}_{p}^{{\text{(F)}}}(x):={f}^*_{\widehat{\bftheta}_n,p}(x), \quad \widehat{g}_{p}^{{\text{(F)}}}(x):=g^*_{\widehat{\bftheta}_n, p}(x),
$$
where the superscript \lq\lq$\text{(F)}$" stands for \lq \lq frequentist" and $\widehat{\bftheta}_n=(\widehat{\sigma}_n, \widehat{\gamma}_n)$.
By means of Corollary \ref{cor:newapprox} the accuracy of  the above estimators can be assessed by quantifying their rates of contraction to  the true densities $f_{t'}^*$ and $g_{t'}^*$ in Hellinger distance. This is formally stated by the next result.
\begin{prop}\label{pro:rate_Helli}
Assume that the conditions of Corollary \ref{cor:newapprox} are satisfied. Assume also that the following conditions are satisfied as $n\to \infty$:
\begin{inparaenum}\label{prop:cons}
\item\label{cond:k} $k\to \infty$ as $n\to\infty$ and $k=o(n)$,
\item\label{cond:balance} $\sqrt{k}|A(n/k)|\to \lambda \in (0,\infty) $,
\item\label{cond:est} 
$|\widehat{\gamma}_n-\gamma|=O_{\mathbb{P}}(1/\sqrt{k})$ and $ |\widehat{\sigma}_n/s(t)-1|=O_{\mathbb{P}}(1/\sqrt{k})$.
\end{inparaenum}
Finally, let $z_{\gamma}(x)=w_{\gamma}(x)$, if $\gamma\geq 0$, and $z_{\gamma}(x)=\log(x)w_{\gamma}(x)$, if $\gamma < 0$.
Then, for any sequence $\epsilon_n \to 0$ and such that $k\epsilon_n^2 \to \infty$ as $n\to\infty$ we have:
	\begin{inparaenum}
	\item[(i)] If $ k/(np) \to \nu  \in [1,\infty)  $ as $n \to \infty$, 
	$$
	\hell(f_{t'}^*, 
	{\widehat{f}}_{p}^{{\text{(F)}}}
	)=O_{\mathbb{P}}\left(\epsilon_n\right), \quad 
	\hell(g_{t'}^*, 
	\widehat{g}_{p}^{{\text{(F)}}}
	)=O_{\mathbb{P}}\left(\sqrt{\epsilon_n} \right).
	$$
	\item[(ii)] If instead $k/(np) \to \infty$ and $ -\epsilon_n \log(np/k) \to 0$ as $n\to\infty$,
	$$
	\hell(f_{t'}^*, 
	{\widehat{f}}_{p}^{{\text{(F)}}}
	)=O_{\mathbb{P}}\left( -\epsilon_n \log(np/k)  \right).
	$$
	\item[(iii)] Moreover, if  $k/(np) \to \infty$ and $\epsilon_n z_{\gamma}\left(k/(np)\right) \to 0$ as $n\to\infty$,
	$$\hell(g_{t'}^*, 
	\widehat{g}_{p}^{{\text{(F)}}}
	)=O_{\mathbb{P}}\left(
	\sqrt{\epsilon_n z_{\gamma_0}\left(\frac{k}{np}\right)}
	 \right).
	$$
\end{inparaenum}
\end{prop}
Assumptions \ref{cond:k}--\ref{cond:balance} of Proposition \ref{prop:cons} have been widely used in the literature to show that standard estimators $\widehat{\gamma}_n$ and $\widehat{\sigma}_n$ of the true values $\gamma$ and $\sigma$, proposed within the POT approach  \citep[see e.g.][Ch. 3--5]{dehaan+f06}, satisfy under the second order condition the asymptotic normality of the sequence
$$
\sqrt{k}\left(
\widehat{\gamma}_n-\gamma,
\frac{\widehat{\sigma}_n}{s(U(n/k))}-1
\right).
$$ 
Therefore, such estimators also comply with the assumption \ref{cond:est} of Proposition \ref{prop:cons},  whose statement allows to readily obtain the rate of contraction of ${\widehat{f}}_{p}^{{\text{(F)}}}$ and $\widehat{g}_{p}^{{\text{(F)}}}$ to $f_{t'}^*$ and $g_{t'}^*$  in Hellinger distance. 
We provide two examples of classical estimators that satisfy the conditions of Proposition \ref{pro:rate_Helli} and comply with its results.
\begin{ex}\label{ex:MLE}
Under the assumptions of Proposition \ref{theo:hellrate} and conditions \ref{cond:k}--\ref{cond:balance} of Proposition \ref{prop:cons}, with probability tending to 1 there exists a unique sequence of ML estimators of $\gamma$ and $\sigma$ given by
$$
	(\widehat{\sigma}_n, \widehat{\gamma}_n)= \underset{ \bftheta \in\Theta}{\arg\max} \prod_{i=1}^{k}h_\bftheta \left(
	X_{n-k+i,n}-X_{n-k,n}
	\right)
	$$
	where $\Theta=(0,\infty)\times(-1/2,\infty)$, satisfying condition \ref{cond:est} of Proposition \ref{prop:cons}, see \cite{drees04, zhou09} and \citet[][Corollary 2.3]{dombry23}.
\end{ex}
\begin{ex}\label{ex:GPWM}
	The GPWM estimators of $\gamma$ and $\sigma$ are defined as
	$$
	\widehat{\gamma}_n=1-\left(\frac{P_n}{2 Q_n}-1\right)^{-1}, \quad \widehat{\sigma}_n=P_n\left(\frac{P_n}{2 Q_n}-1\right)^{-1},
	$$
	where
	$$
	P_n=\frac{1}{k}\sum_{i=1}^{k}\left(X_{n-i+1,n}-X_{n-k,n}\right), \quad Q_n=\frac{1}{k}\sum_{i=1}^{k}\frac{i}{k}\left(X_{n-i+1,n}-X_{n-k,n}\right).
	$$
	Under the assumptions of Proposition \ref{theo:hellrate} and conditions \ref{cond:k}--\ref{cond:balance} of Proposition \ref{prop:cons}, and assuming further that $\gamma<1/2$, such estimators
	satisfy condition \ref{cond:est} of Proposition \ref{prop:cons}, see e.g. Theorem 3.6.1 in \cite{dehaan+f06}.
\end{ex}
Exploiting the results of Proposition \ref{pro:rate_Helli}, next corollary establishes the accuracy of predictive regions for future peaks variables {above the threshold $t'=Q(p)$.}
%{\color{magenta}above an extreme threshold $t'=U_0(1/\tau)$.}
%
\begin{cor}\label{cor:freqpred}
Assume that conditions of Proposition \ref{pro:rate_Helli} are satisfied. Let $\mathcal{P}_{p}$ be a measurable set depending on sequence of excesses {$X_{n-k+i,n}-X_{n-k,n}$, $i=1,\ldots,k$,} %representing an extreme region  
and satisfying for $\alpha \in (0,1)$,
$$
\int_{\mathcal{P}_{p}} \widehat{g}_{p}^{\text{(F)}}(x) \diff x =1-\alpha.
$$
For any sequence $\epsilon_n \to0$ and such that $n \epsilon_n^2 \to \infty$ as $n \to \infty$ we have:
 \begin{inparaenum}
 \item[(i)]
 $
 \Prob \left(
 X  \in \mathcal{P}_{p} | X > t'
 \right)= 1-\alpha + O_{\Prob}(\sqrt{\epsilon_n}),
 $
 if $ k/(np) \to \nu  \in [1,\infty)  $ as $n \to \infty$.
 \item[(ii)] 
 $
 \Prob \left(
 X  \in \mathcal{P}_{p} | X > t'
 \right)= 1-\alpha + O_{\Prob}(
 \sqrt{
 \epsilon_n z_{\gamma} \left(k/(np)\right)
 }),
 $
 if $k/(np) \to \infty$ and $\epsilon_n z_{\gamma_0}\left(k/(np)\right)$ $\to 0$, as $n\to\infty$.
 \end{inparaenum}
\end{cor}
A concrete example of how to select $k$, $p$ and $\epsilon_n$ so that the results in Corollary \ref{cor:freqpred} hold is reported next.
\begin{ex}\label{ex:fredpred}
Set $k=n^\delta \log^\eta(n)$, with $\delta \in (0,1)$, $\eta \in \Real$, and $p= k^\zeta/n$, with $\zeta \in (0,1)$. Set $\epsilon_n=C_n/\sqrt{k}$, with $C_n$ being a sequence going to infinity arbitrarily slowly.  Then, Corollary \ref{cor:freqpred}(ii) guarantees {that} the prediction of peaks  over the true extreme quantile $Q(p)$ by means of the extreme regions $\mathcal{P}_{p}$, deduced 
%by
{from} 
the predictive density $\widehat{g}_{p}^{\text{(F)}}$, implies a probability error that is not larger than $n^{-\delta \zeta/4}$, up to a logarithmic term proportional to $\sqrt{C_n} (\log n)^{1/2-(\eta \zeta)/4}$.
\end{ex}

\subsection{Bayesian approach}\label{sec:bayes}
A Bayesian procedure {for} inference with the POT method that includes 
%the 
classical {prior} formulation and also empirical Bayes {ones} is synthesized as follows \citep[see][for a complete description]{dombry23}.
The {considered}
%proposed 
procedure relies on a flexible definition of a prior distribution %regarding 
{for}
the GP distribution parameter $\bftheta $, with its density function of the  form
\begin{equation}\label{eq:prior_ist}
	\pi(\bftheta)= \pi_{\text{sh}}(\gamma) \pi_{\text{sc}}^{(n)}
	\left( \sigma\right), \quad \bftheta \in \Theta,
\end{equation}
where $ \pi_{\text{sh}}$ is a prior density on $\gamma$ and for each $n=1,2,\ldots$, $\pi_{\text{sc}}^{(n)}$ is a prior density on $\sigma$, whose expression may or may not depend on $n$, and where $\Theta$ is as in Example \ref{ex:MLE}.
{In order for}
	%For the purpose of 
	the corresponding posterior distribution {to} fulfil some desirable features, the following properties are assumed.
\begin{cond}\label{cond:prior}
	The densities  $\pi_{\text{sh}}$ and $\pi_{\text{sc}}^{(n)}$ are such that:
	\begin{inparaenum}
		%
		%%%%%%%%%%%
		\item\label{cond:pisc} For each $n=1,2,\ldots$, $\pi_{\text{sc}}^{(n)}:\Real_+\to\Real_+$ and given $t=Q(k/n)$
		\begin{inparaenum}
			\item[(a.1)]\label{posit} there is $\delta>0$ such that $\pi_{\text{sc}}^{(n)}(s(t))s(t)>\delta$ and for any $\eta>0$ there is $\epsilon>0$ such that for all $\sigma\in(1\pm\epsilon)$ 
			$$
			\pi_{\text{sc}}^{(n)}(s(t)\sigma)\in(1\pm\eta)\pi_{\text{sc}}^{(n)}(s(t));
			$$
			\item[(a.2)]\label{piscbound} there is $C>0$ such that  
			$\pi_{\text{sc}}^{(n)}(s(t)\sigma)\leq C/(\sigma s(t))${.}
		\end{inparaenum}
		If $\pi_{\text{sc}}^{(n)}$ is data-dependent then (a.1) and (a.2) hold with probability tending to $1$.
		\item\label{cond:pish}  $\pi_{\text{sh}}$ is a positive and continuous function at the true value {of} $\gamma$ satisfying $\int_{-1/2}^0 \pi_{\text{sh}}(\gamma)\,\mathrm{d}\gamma<\infty$ and $\sup_{\gamma>0}\pi_{\text{sh}}(\gamma)<\infty$.
	\end{inparaenum}
\end{cond}
For a prior density $\pi$ as in \eqref{eq:prior_ist}, the posterior distribution of the GP distribution parameter $\bftheta$ is given by Bayes rule
\begin{equation}\label{eq:posterior_pot}
	\Pi_n(B)=\frac{\int_{B}\prod_{i=1}^k h_\bftheta(X_{n-i+1,n}-X_{n-k,n})\pi(\bftheta)\diff \bftheta}
	{\int_{\Theta}{\prod_{i=1}^k h_\bftheta(X_{n-i+1,n}-X_{n-k,n})}\pi(\bftheta)\diff \bftheta},
\end{equation}
for all measurable sets $B\subset \Theta$. According to \citet[][Theorem 2.10]{dombry23}, as $n$ increases, the posterior distribution $\Pi_n$ concentrates on the value{s} $(\gamma,\sigma)\in\Theta$ such that $|\gamma -\gamma_0|$ and $|\sigma/s_0(t) -1|$ are {close} to $0$. However, the interesting issue of assessing the accuracy of Bayesian density estimation within the POT method has been left untouched so far. Such a gap is filled by the following result which guarantee{s} that the posterior distribution $\Pi_n$ asymptotically concentrates on values $\bftheta\in\Theta$ such that the corresponding GP densities ${f}_{\bftheta,{p}}^{*}$ and  $g_{\bftheta,{p}}^{*}$ are {close} to the true predictive densities $f_{t'}^*$ and $g_{t'}^*$  of the excess variable and the peak variable, respectively, in Hellinger distance.
\begin{prop}\label{prop:hellcons}
Assume that Condition \ref{cond:prior}, the conditions of Corollary \ref{cor:newapprox} and conditions \ref{cond:k}--\ref{cond:balance} of Proposition \ref{prop:cons} are satisfied. Then, for any sequence $\epsilon_n \to0$ such that  $n \epsilon_n^2 \to \infty$, as $n \to \infty$,  there is a $R>0$ such that:
	\begin{inparaenum}
	\item[(i)] If $ k/(np) \to \nu  \in [1,\infty)  $ as $n \to \infty$, 	
	\begin{inparaenum}
	\item[(i.a)]
	$
	\Pi_n\left(
		\left\lbrace
		\bftheta \in \Theta: \, \hell ({f}_{\bftheta,{p}}^{*},f_{t'}^*) > \epsilon_n
		\right\rbrace
		\right) = O_{\mathbb{P}}\left(
		e^{-R k \epsilon_n^2 }
		\right);
	$
	\item[(i.b)]
	$	\Pi_n\left(
		\left\lbrace
		\bftheta \in \Theta: \, \hell (g_{\bftheta,{p}}^{*},g_{t'}^*) > \sqrt{\epsilon_n}
		\right\rbrace
		\right) = O_{\mathbb{P}}\left(
		e^{-R k \epsilon_n^2 }
		\right).
	$
	\end{inparaenum}
	\item[(ii)] If  $k/(np)\to \infty$ and $-\epsilon_n \log(np/k) \to 0$ as $n \to \infty$,
	%	\begin{inparaenum}
	%	\item[(i.a)]
		$$
		\Pi_n\left(
		\left\lbrace
		\bftheta \in \Theta: \, \hell ({f}_{\bftheta,{p}}^{*},f_{t'}^*) > -\epsilon_n \log(np/k)
		\right\rbrace
		\right) = O_{\mathbb{P}}\left(
		e^{-R k \epsilon_n^2}
		\right).
		$$
		\item[(iii.)] Finally, if $k/(np)\to \infty$ and $\epsilon_n z_{\gamma}\left(k/(np)\right) \to 0$ as $n \to \infty$,
		$$	
		\Pi_n\left(
		\left\lbrace
		\bftheta \in \Theta: \, \hell (g_{\bftheta,{p}}^{*},g_{t'}^*) > \sqrt{\epsilon_n z_{\gamma}\left(k/(np)\right)}
		\right\rbrace
		\right) = O_{\mathbb{P}}\left(
		e^{-R  k \epsilon_n^2 }
		\right).
		$$
	%\end{inparaenum}
\end{inparaenum}
\end{prop}
Under the Bayesian paradigm, an estimator of the true predictive density $f_{t'}^*$ of a future unobservable excess variable over the true high threshold $t'=Q(p)$, with $p\leq k/n$, is given by
$$
\widehat{f}_{{p}}^{{\text{(B)}}}(x) := \int_{\Theta} {f}_{\bftheta,{p}}^{*}(x) \diff \Pi_n (\bftheta),
$$
where the superscript \lq \lq $\text{(B)}$" stands for \lq \lq Bayesian". 
%The benefit of such a Bayesian estimator is that is optimal under square loss of the true density $f_{t'}^*$.
%
{Similarly,} an estimator of the true predictive density $g_{t'}^*$ of a future unobservable peak variable is given by
\begin{equation}\label{eq:post_pred_dens}
\widehat{g}_{{p}}^{{\text{(B)}}}(x) := \int_{\Theta} g_{\bftheta,{p}}^{*}(x) \diff \Pi_n (\bftheta).
\end{equation}
Next result establishe{s} the accuracy of such an estimator $\widehat{g}_{{p}}^{{\text{(B)}}}$  and that of predictive region  $\mathcal{P}_{p}$ derived from it. 
\begin{cor}\label{cor:bayespred}
Assume that conditions of Proposition \ref{prop:hellcons} are satisfied. Let $\mathcal{P}_p$ be a measurable set depending on sequence of excesses $X_{n-k+i,n}-X_{n-k,n}$, $i=1,\ldots,k$, 
%{\color{magenta}representing an extreme region} and 
{and} satisfying for $\alpha \in (0,1)$
$$
\int_{\mathcal{P}_p} \widehat{g}_{{p}}^{\text{(B)}}(x) \diff x =1-\alpha.
$$ 
For any sequence $\epsilon_n\to 0$ as $n \to \infty$ we have: 
\begin{inparaenum}
\item[(i)] If $ k/(np) \to \nu  \in [1,\infty) $ as $n \to \infty$ and $\log k=o(n \epsilon_n^2)$, then 
\begin{inparaenum}
\item[(i.a)] 
$\hell(\widehat{f}_{{p}}^{{\text{(B)}}}, f_{t'}^*)=O_{\Prob}\left(\epsilon_n\right) $;
\item[(i.b)] 
$\hell(\widehat{g}_{{p}}^{{\text{(B)}}}, g_{t'}^*)=O_{\mathbb{P}}\left(\sqrt{\epsilon_n} \right)${;} 
\item[(i.c)]  
$
\Prob\left(X  \in \mathcal{P}_{p} | X > t'\right)= 1-\alpha + O_{\Prob}(\sqrt{\epsilon_n}).
$
\end{inparaenum}
\item[(ii)] If $k/(np) \to \infty$, $-\epsilon_n \log(np/k) \to 0$ as $n \to \infty$, $\log( k \log^2(np/k) ) = o(k \epsilon_n^2)$, then
$$\hell(\widehat{f}_{{p}}^{{\text{(B)}}}, f_{t'}^*){=O_{\mathbb{P}}(-\epsilon_n \log(np/k) )}.$$
\item[(iii)] If $k/(np) \to \infty$, $\epsilon_nz_{\gamma}\left(k/(np)\right) \to 0$ as $n\to\infty$, $\log\left(k z_{\gamma}^2 \left(k/(np)\right) \right)=o(k\epsilon_n^2)$, then 
$$
\hell(\widehat{g}_{{p}}^{{\text{(B)}}}, g_{t'}^*)=O_{\Prob}\left( \sqrt{\epsilon_nz_{\gamma}\left(k/(np)\right)}\right)
$$
and
$$
\Prob(X\in\mathcal{P}_{p} | X > t')= 1-\alpha + O_{\Prob}\left(\sqrt{\epsilon_nz_{\gamma}(k/(np}\right).
$$
\end{inparaenum}
\end{cor}
Note that, similar{ly} to the frequentist case, with {e.g.}  the same selection of  $k$, $p$ and $\epsilon_n$ as in Example \ref{ex:fredpred}, Corollary \ref{cor:bayespred}(iii) also guarantees that prediction based on the extreme region $\mathcal{P}_{p}$ deduced by the posterior predictive density $\widehat{g}^{{\text{(B)}}}_{p}$ implies a small probability error in such a forecasting task.

\section{Milan extreme temperatures analysis}\label{sec:real_analysiss}

We analyse the summer temperatures of Milan in Italy from $1991$ to $2023$. We consider the Daily Maximum Temperatures (DMT) recorded between June to September. For the purpose of using an accessible forecasting technique we assume that DMT are independent, although they are temporally dependent in practice. Since we focus only on summer temperatures, however, there is no seasonal effect in the data.  In the summer of 2003 one of the most massive heat waves of the last decades hit Europe with temperature records broken in several European cities.  In Milano on August 11th the hottest temperature ever recorded before reached $38.3$ °C . Temperatures were particularly unbearable by the high level of humidity in the air, which is a typical feature of big urban cities as Milan. This analysis {aims} to predict temperature peaks over certain thresholds and accordingly understand what is the threshold beyond which a future peak temperature like the record one or even higher is plausible to be reached.

Once the missing values {are removed}, $n=3140$ DMT are available. A high threshold equal to $t=34$ °C is set using the order statistic $X_{n-k,n}$, with $k=169$, so that there are only $k/n\approx 5.4\%$ of hottest temperatures in the sample. The excess temperatures are used to fit the GP distribution using the MLE, GPWM \citep[e.g.][Ch. 3]{dehaan+f06} and the Bayesian approach \citep{dombry23}, where a data dependent prior distribution for $\sigma$ and a data independent one for $\gamma$ is used.  

\begin{figure}[t!]
	\centering
	\includegraphics[width=0.27\textwidth, page=3]{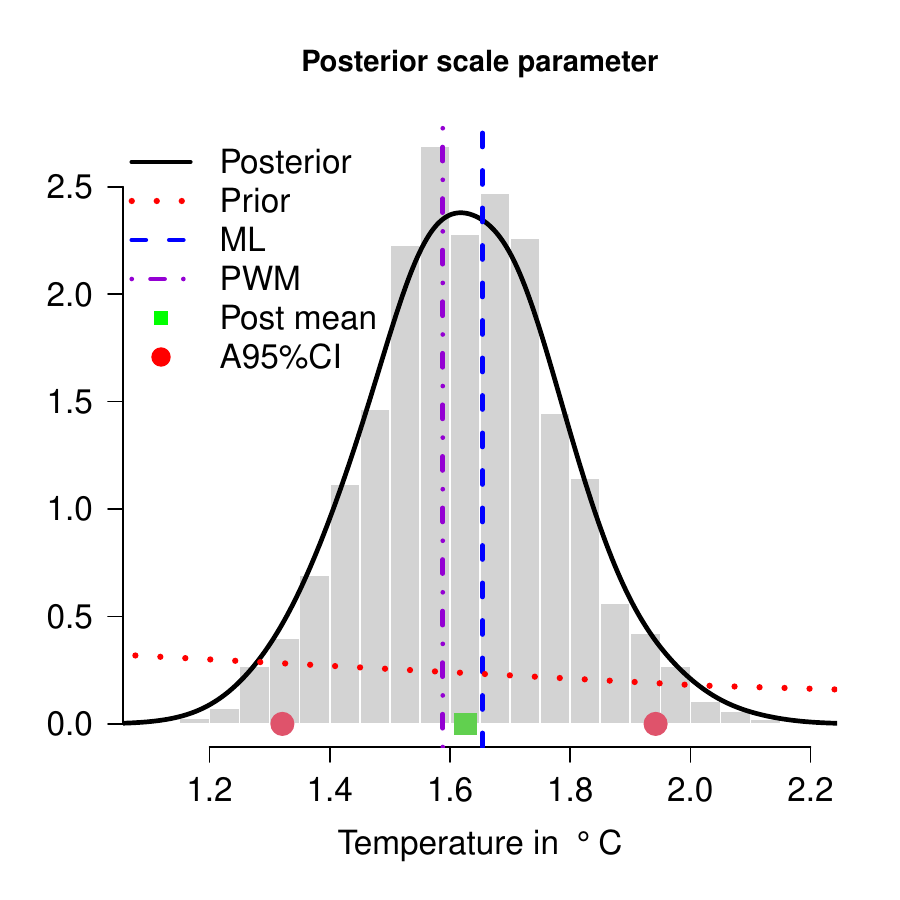}
	\includegraphics[width=0.27\textwidth, page=4]{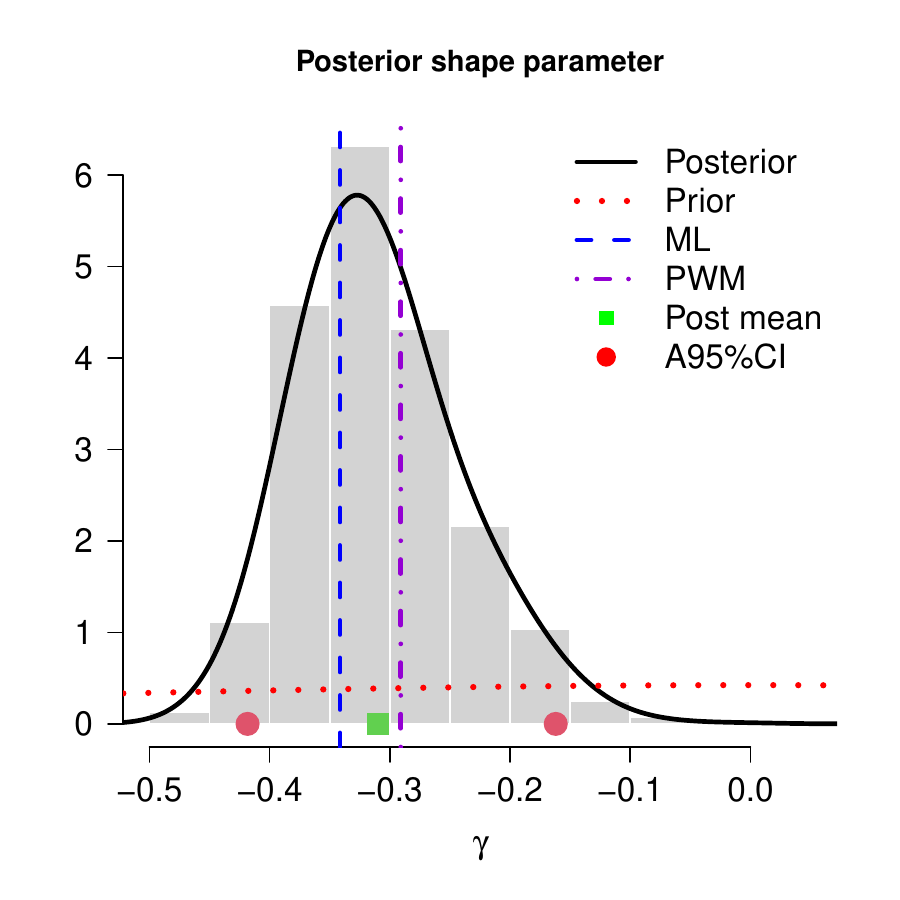}
	\includegraphics[width=0.27\textwidth, page=5]{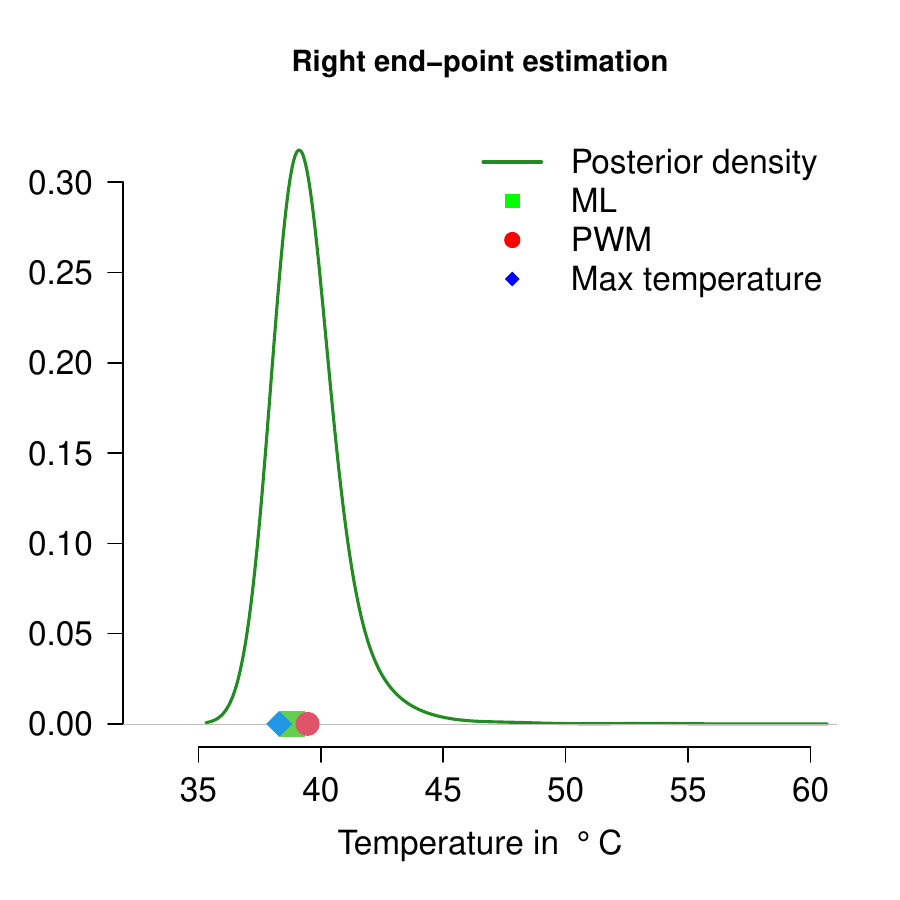}\\
	\includegraphics[width=0.27\textwidth, page=7]{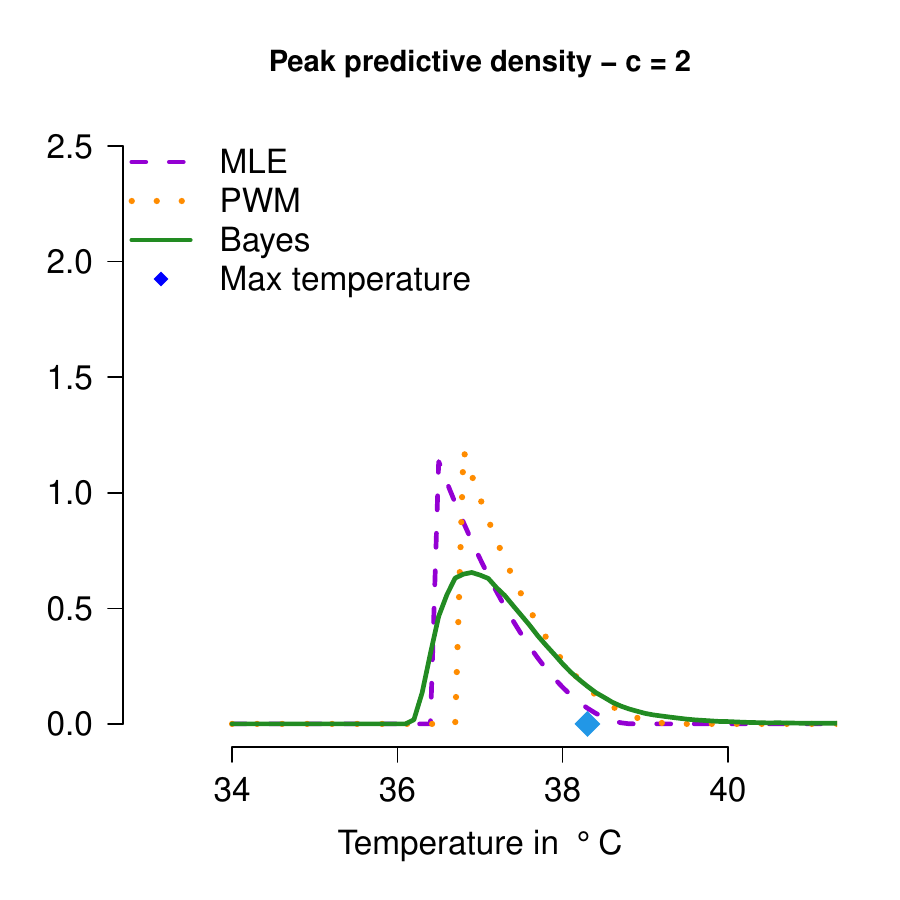}
	\includegraphics[width=0.27\textwidth, page=8]{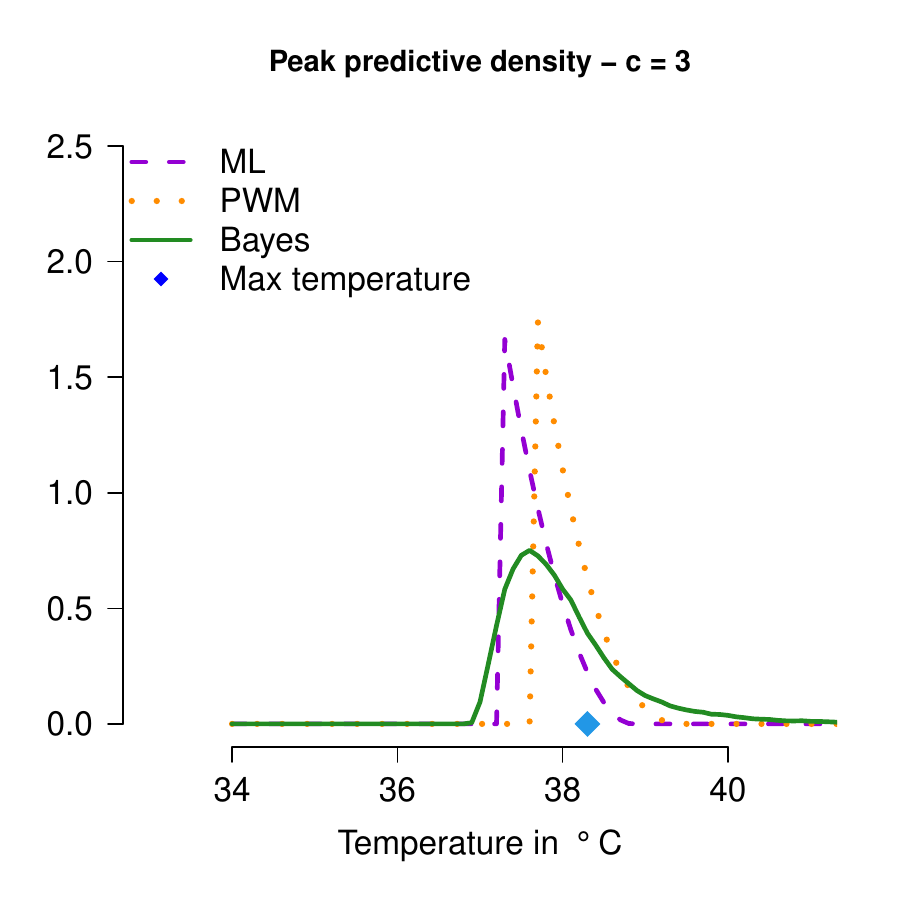}
	\includegraphics[width=0.27\textwidth, page=9]{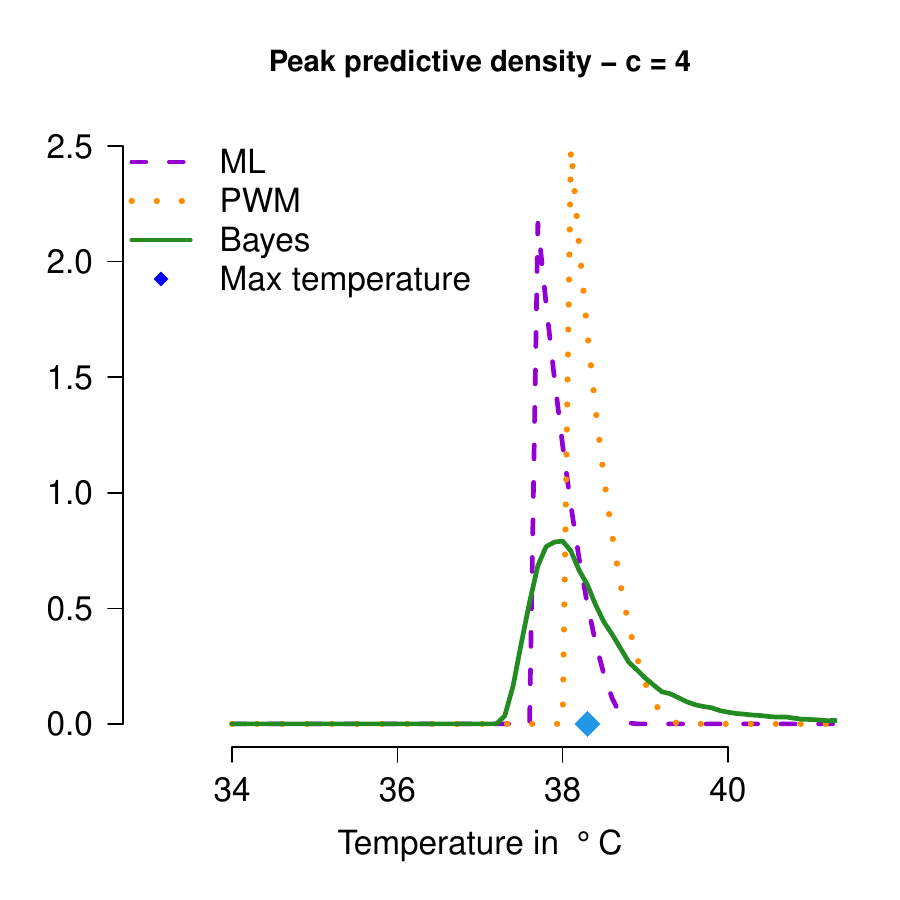}
	\caption{Estimation results. Posterior (black solid line) and prior (red dotted line) densities of the scale parameter (top-left panel) and extreme index (top-middle panel). The green squares and red circles are the posterior mean and lower and upper bounds of the Asymmetric  95\% Credibility Interval (A95\%CI). Vertical blue dashed and violet dot-dashed lines ate the ML and GPWM estimates of such parameters. Posterior density of the right end-point (top-right panel) and its ML and GPWM estimates. Estimated predictive densities (bottom panels) of peaks over extreme thresholds obtained with  $c=2,3,4$ (from left to right) and the ML (violet dashed line), GPWM (orange dotted line) and Bayesian methods (green solid line).}\label{fig:real_analysis}
\end{figure}

The ML and GPWM estimates of $(\sigma, \gamma)$ are $(1.65, -0.34)$ and $(1.59, -0.29)$, respectively. Top-left and top-middle panels of Figure \ref{fig:real_analysis} show the empirical posterior densities of such parameters, obtained with a sample of $M=20$th values drawn from the posterior \citep[see][for details on the MCMC algorithm used]{padoan24b}, whose means are $(1.63, -0.31)$ and asymmetric $95\%$ credibility intervals 
%for $(\sigma, \gamma)$ 
are $[1.32, 1.94]$ and $[-0.42, -0.16]$, respectively, derived using the posterior quantiles. These results suggest that the distribution $F$ of summer DMT is short-tailed with a finite upper end-point $x^*$, as expected. 
We then estimate $x^*$, to understand how far temperatures hotter than those observed 
%with 
{in} the sample could plausibly rise in the future.  Its ML and GPWM estimates are $38.84$ °C and $39.46$ °C and the top-right panel of Figure \ref{fig:real_analysis} displays its empirical posterior density, whose mean is 39.46 °C and the corresponding  asymmetric $95\%$ credibility interval is $[38.43$ °C, $42.54$ °C$]$. 
\begin{table}[t!]
	\centering
	\caption{Estimation results. Asymmetric 95\% Predictive Intervals (A95\%PI) obtained with $t=Q(k/n)$ approximated by $X_{n-k,n}$  (4th row) and with the ML (left part), GPWM (middle part) and Bayesian (right part) methods. A95\%PI obtained with $t'=Q(p)$ (3rd, 6th and 9th column and from 6th to the 8th row), where $p$ and $Q(p)$ are approximated by the ML (1st and 2nd column) and GPWM (4th and 5th column) estimates $\widehat{p}$ and $\widehat{Q}(\widehat{p})$ and by a posterior sample from which A95\%CI  are derived (7th and 8th column).}
	{\scriptsize
		\begin{tabular}{ccc|ccc|ccc}
			\toprule
			\multicolumn{9}{c}{Method}\\
			\midrule
			\multicolumn{3}{c}{ML} & \multicolumn{3}{c}{GPWM} &\multicolumn{3}{c}{Bayes}\\
			\midrule
			$k/n$ & $X_{n-k,n}$ & A95\%PI & $k/n$ & $X_{n-k,n}$ & A95\%PI & $k/n$ & $X_{n-k,n}$ & A95\%PI\\
			\midrule
			5.382  & 34.0 & [34.1, 37.5] & 5.382 & 34.0 & [34.1, 37.6] & 5.382 & 34.0 & [34.1, 37.6]\\
			\midrule
			$\widehat{p}$ & $\widehat{Q}(\widehat{p})$ & A95\%PI & $\widehat{p}$ & $\widehat{Q}(\widehat{p})$ & A95\%PI & A95\%CI-$p$ & A95\%CI-$Q(p)$ & A95\%PI\\
			\midrule
			0.707 & 36.4 & [36.4, 37.9] & 0.497 & 36.7 & [36.7, 38.3] & [0.075, 1.027] & [36.2, 38.2] & [36.4, 39.1]\\ 
			0.216 & 37.2 & [37.2, 38.2] & 0.123 & 37.6 & [37.6, 38.7] & [0.006, 0.390] & [36.9, 39.8] & [37.0, 40.2]\\
			0.093 & 37.6 & [37.6, 38.4] & 0.046 & 38.1 & [38.3, 38.8] & [0.001, 0.196] & [37.2, 40.5] & [37.4, 41.6]\\
			\bottomrule
		\end{tabular}
	}
	\label{tab:_real_analysis}
\end{table}
%

%Alternatively to the estimation of a distribution's  parameter, 
Now, we consider the proper probabilistic forecasting of peaks over certain very high thresholds, described by the method of Section \ref{sec:strong_results}. %to understand the likelihood of future temperature peaks and what their temperature rise is plausible.
%
%In addition, we consider probabilistic forecasting of peaks over an extreme threshold, according the approach described in Section \ref{sec:strong_results}, to predict what temperature rise is plausible in the future. 
%
To start, we consider the predictive density of a peak over the initial threshold $t=34$ °C, which is obtained setting $p=k/n$ in the formula of $g_{\bftheta,{p}}^*$ in \eqref{eq:predictive_dist}.  Frequentist estimates of it are obtained plugging-in the ML and GPWM estimates of $(\sigma, \gamma)$ in $g_{\bftheta,{p}}^*$. A Bayesian estimate is obtained instead computing the posterior predictive density in \eqref{eq:post_pred_dens}, which we approximate according to the Monte Carlo approximation
$$
\widehat{g}^{{\text{(B)}}}(x) = \sum_{i=1}^M h^*_{\widetilde{\bftheta}_i, {p}}\left(x-X_{n-k,n} + \widetilde{\sigma}_i \frac{\left(\frac{np}{k}\right)^{-\widetilde{\gamma}_i}-1}{\widetilde{\gamma}_i}\right),
$$
where $\widetilde{\bftheta}_i=(\widetilde{\sigma}_i, \widetilde{\gamma}_i)$, with $i=1,\ldots,M$, is a sample from the posterior of the GP parameters. When ${p}=k/n$ the formula clearly simplifies accordingly. The fourth row of Table \ref{tab:_real_analysis} reports asymmetric 95\% predictive intervals derived from such estimated densities. %estimated with the frequentist and Bayesian methods. 
The considered threshold is not so extreme, it is not surprising then to see that the three methods propose the same interval, which suggests that if we know that a temperature is above 34 °C (we expect this to happen approximately 5.4\% of the time) then it is plausible that it is between approximately 34.1 °C and 37.6 °C.

To safeguard against even severer events, the predictive density of future peaks above even higher thresholds is estimated. The following approach explains how to select a very high threshold according to a guideline of easy interpretation.  With a short-tailed distribution $F$, with $\gamma<0$, we know that for {any} 
%$v>1$ and 
$x>0$ we have $(x^*-Q(1/(xv)))/(x^*-Q(1/v)) \to x^\gamma$ as $v\to {\infty}
%x^*
$ \citep[][Theorem 1.1.13]{dehaan+f06}. Now, 
%if you set
{setting} 
$v=n/k$ and $x=k/(np)$, where $p\leq k/n$ is a small level used to compute a very high threshold, then such a 
condition 
reads as $(x^*-Q(p))\approx (k/(np))^\gamma (x^*-Q(k/n))$ as $n\to\infty$. If you further set $c=(k/(np))^{-\gamma}$, then the condition finally reads as $(x^*-Q(p))\approx (x^*-Q(k/n))/c$, as $n\to\infty$. This  means that we can set a scaling factor $c>0$ such that the gap between the right end-point  $x^*$ of $F$ and the very high threshold $t'=Q(p)$ is equal for instance to half ($c=2$), a third ($c=3$), etc., of the gap between $x^*$ and the initial threshold $t=Q(k/n)$. We specify $c=2,3,4$ and then estimate $p=c^{1/\gamma} k/ n$ and the very high threshold $Q(p)$.

In the last 3 lines of Table \ref{tab:_real_analysis} are reported $\widehat{p}=c^{1/\widehat{\gamma}} k/ n$ in percentage and $\widehat{Q}(\widehat{p})=X_{n-k,n}+\widehat{\sigma}(({n\widehat{p}/k})^{-\widehat{\gamma}}-1)/\widehat{\gamma}$ in °C, where the estimates $(\widehat{\sigma},\widehat{\gamma})$ are obtained with the ML (1st and 2nd column) and GPWM (4th and 5th column) methods and asymmetric $95\%$ credibility intervals for $p$ and $Q(p)$ (7th and 8th column), obtained with the posterior quantiles of $\widetilde{p}_i=c^{1/\widetilde{\gamma}_i} k/ n$ and  $\widetilde{Q}_i(\widetilde{p}_i)=X_{n-k,n}+\widetilde{\sigma}_i(({n\widetilde{p}_i/k})^{-\widetilde{\gamma}_i}-1)/\widetilde{\gamma}_i$, where again $(\widetilde{\sigma}_i, \widetilde{\gamma}_i)$, with $i=1,\ldots,M$, is a posterior sample of the GP parameters.
The thresholds estimated with the GPWM method are higher than those obtained with the ML one, and their gap increase with the increasing of $c$. A more extended range of plausible values is instead suggested {by} the Bayesian approach.
The bottom panels of Figure \ref{fig:real_analysis} display the corresponding estimated predictive densities, obtained with the ML (yellow dotted lines) and GPWM (violet dashed lines) methods and the Bayesian approach (green solid lines). Consistently with the previous findings, the dissimilarity between the predictive densities increases with the increasing of $c$, with those estimated with GPWM method that place gradually  more mass on hotter temperatures than those obtained with ML method. The Bayesian approach seems to be much better balanced suggesting wider predictive densities, allowing for both lower and hotter temperatures. Table \ref{tab:_real_analysis} also reports along the last 3 rows the corresponding asymmetric 95\% predictive intervals obtained with the ML (3rd column), GPWM (6th column) and Bayesian (9th column) methods. The latter approach suggests that if we know that a temperature is above a threshold between 36.2 °C (36.9 °C or 37.2 °C) and 38.2 °C (39.8 °C or 40.5 °C), we expect this to happen between 0.075\% (0.006\% or 0.001\%) and 1.027\% (0.390\% or 0.196\%) of the time, then it is plausible that the temperature ranges between 36.4 °C (37.0 °C or 37.4 °C) and 39.1 °C (40.2 °C or 41.6 °C). Such a range indeed includes a hot temperature as the record one or even hotter. 
%Similar interpretations can be achieved 
{The results obtained} with the other two estimation methods {can be interpreted in a similar way}. 

\section{Conclusion}\label{sec:conclusion}
Probabilistic forecasting of future excesses or peaks over a high threshold, for several threshold levels corresponding to different degrees of extrapolation beyond the range of observed data, has been accurately investigated in this paper. Our frequentist and Bayesian proposed tools for performing probabilistic forecasting of future excesses or peaks, are mathematical grounded, simple to implement and easy to interpret for practical applications. To guarantee that the proposed forecasting tools, based on predictive densities, are easily interpretable in the real applications and a rigorous study of their asymptotic behaviour is feasible, we relied on the assumption of independent data and independence between also the later and events to be predicted.

Since in many relevant applications the data are time dependent is then particularly important to extend our results in the case that one works with a stationary sequence of serially dependent observations, in order to improve the prediction further. However, this is a very ambitious objective, since it requires a full comprehension of several still open problems. These include, e.g., establishing local and global asymptotic behaviour of the GP likelihood function, and understanding how to integrate the information on the serial dependence in order to perform prediction at successive time lags in an interpretable way, etc. These are challenges for future research.
\section*{Acknowledgements}
Simone Padoan is supported by the Bocconi Institute for Data Science and Analytics (BIDSA), Italy, thanks also to MUR - Prin 2022 - Prot. 20227YZ9JK.

\bibliographystyle{chicago} 
\bibliography{bibliopm_final3}

\end{document}